\documentclass[aps,prb,reprint]{revtex4-1}
\usepackage[version=3]{mhchem} 
\usepackage[T1]{fontenc}       
\usepackage{graphicx}
\usepackage{bm}
\usepackage{xcolor}
\usepackage{lineno}
\usepackage{overpic}
\usepackage[final]{changes}

\begin{document}
\title{Dynamics and ordering of weakly Brownian particles in directional drying}

\author{Cecile Noirjean}
\email{ce.noirjean@gmail.com}
\author{Moreno Marcellini}
\affiliation{Ceramic Synthesis and Functionalization Lab, UMR3080 CNRS/Saint-Gobain, 84306 Cavaillon, France}

\author{Thomas E. Kodger}
\affiliation{Physical Chemistry and Soft Matter, Wageningen University \& Research, Stippeneng 4, 6708WE Wageningen, The Netherlands}

\author{C\'{e}cile Monteux}
\affiliation{SIMM, UMR 7615 CNRS-ESPCI-Universit\'e Pierre et Marie Curie, ESPCI, Paris, France}

\author{Sylvain Deville}
\email{sylvain.deville@saint-gobain.com}
\affiliation{Ceramic Synthesis and Functionalization Lab, UMR3080 CNRS/Saint-Gobain, 84306 Cavaillon, France}

\date{\today}

\begin{abstract}
Drying of particle suspensions is an ubiquitous phenomenon with many natural and practical applications. \added{In particular,} in unidirectional drying, the evaporation of the solvent induces flows which accumulate particles at the liquid/air interface. The progressive build-up of a dense region of particles can be used, in particular, in the processing of advanced materials and architectures while the development of heterogeneities and defects in such systems is critical to their function. 

A lot of attention has thus been paid to correlate the flow and particles dynamics to the ordering of particles. However, dynamic observation at the particle scale and its correlation with local particle ordering are still missing. Here we show \added{by measuring the particle velocities }with high frame rate laser scanning confocal microscopy that the ordering of weakly Brownian particles during \replaced{unidirectional }{directional} drying \added{in a Hele-Shaw cell opened on one side} depends on the \replaced{velocity of particles that impinge at the pinned liquid/solid interface.}{particle velocity.} \added{Under the ambient and experimental conditions presented in the following, the particle velocities accumulate in two branches.} A higher degree of ordering is found for \added{the branch of }faster particle velocity which we explain by an increase in the pressure drop which drags the particles into a denser packing as the flow velocity increases. This counter-intuitive behaviour is the opposite to what is found with Brownian particles, which can reorganize by Brownian motion into denser packing during drying, as long as the flow velocity is not too high. These results show that different \replaced{kinetic}{drying} conditions can be used to obtain dense, defect-free regions of particles after drying. \added{In particular, it }suggests that rapid, directional drying could be used to control the crystallinity of particle deposits.
\end{abstract}
\maketitle

\section{Introduction}
The drying of colloidal or non-colloidal suspensions is a very common phenomenon with many natural and technological occurrences. Drying and its results  are of interest in geophysics (drying of soils),~\cite{Bens:2001} food engineering,~\cite{sadek_2014} paints,~\cite{van_der_kooij_watching_2015,Goeh:2017} and ceramic materials science.~\cite{scherer_theory_1990} In the latter, drying is a critical step of many processing and shaping routes for advanced materials. The structural evolution of a typical suspension during drying is the result of the interplay between many parameters, which makes investigations difficult. Of particular interest are the drying patterns and the organisation of particles in the dried deposit, as the development of heterogeneities and cracks in these deposits is critical for many applications. The processing of defect-free colloidal crystals -- which could be achieved by drying -- is of interest for photonic applications, with a band gap in the visible.~\cite{Blan:2000} Lattice defects, such as planar stacking faults or other heterogeneities, have nevertheless a critical impact on the optical properties of such materials.~\cite{Vlas:2000} Understanding the formation mechanism of these structures and their defects during drying is therefore essential.

Various model experiments have thus been developed to understand the fundamentals of drying. In films or sessile droplets, curved fronts are formed and the control of the evaporation conditions is difficult, which result in a variety of drying and crack patterns.~\cite{Alla:1995,Lido:2014} \added{Loussert \textit{et al.}~\cite{Lous:2016} developed a method to observe convection and diffusion by using Raman microspectroscopy while drying charged nanoparticles dispersions in thin-film droplet confined between two circular plates. Moreover, in this work the authors provide spatially resolved measurements of the colloid concentration during the drying.} Unidirectional drying has been developed as a well-controlled method to investigate the underlying phenomena. In unidirectional drying experiments, the evaporation of the solvent occurs at the open-end of the sample. The solvent flow carries the particles towards the opened solvent/air interface, where they organize into a dense packing~\cite{Deeg:1997,Deeg:2000,Rout:2013,Zian:2015,Zian:2015a,Lava:2016} as the progressive accumulation of particles continues.

Two competing phenomena control the drying: the difference of chemical potentials of water at the drying interface and the external humidity, and the removal mechanism of vapour from the drying interface. The latter is almost never precisely controlled. Local fluctuations of humidity in the air at the sample scale -- which can hardly be avoided -- can have a great impact on the drying behaviour of the system. 

The mechanism that builds up the order/disorder deposit at the evaporation interface of drops has been previously identified in the velocity with which the particles impinge on to the pinned interface,~\cite{marin_order--disorder_2011} and the evaporation rate.~\cite{piroird_role_2016} The common hypothesis is that particles form polycrystalline (colloidal) crystals for slow evaporation rate, where particles have enough time to reorganize by Brownian motion after hitting the solid region. The same behavior was observed in sessile drops with large (micron-size) particles: disordered structures were obtained for rapid particle velocity.~\cite{Asko:2013} 
\added{Nevertheless, as it has been shown by Sch{\"o}pe \textit{et al.},~\cite{Scho:2006,Schoe:2007} the polydispersity of particles can influence kinetics and structure of the obtained deposits by changing the volume fraction at the order/disorder transition.}

\added{The rise of microfluidic chips allows to observe the drying of confined droplets and colloidal dispersions: Ziane \textit{et al.}, in two different works, show that homogeneous drying can be achieved by pervaporation of water through a water-permeable membrane, such that colloidal particles can be concentrated up to the growth of a colloidal crystal in the micro-channels.~\cite{Zian:2015a,Zian:2015} We anticipate that although the size of particles is two orders of magnitude smaller than ours and the geometry completely different, these works show that drying colloidal dispersions of charged hard-spheres in microchannels reveals two different regimes of solidification along the channel. In brief, one can observe a homogeneous concentration dampening crossing the liquid-solid transition toward the close-packing, and the classical growth, observed classically in directional drying, of a water-saturated solid of close-packed particles.~\cite{divry_drying_2016} In a successive work, Laval \textit{et al.},~\cite{Lava:2016} by using similar microfluidic device, microfabricated \textit{in-situ} material with predetermined composition. The process strength of the authors' approach relies on the control of the pervaporation process at the microfluidic scale to build materials ranging from colloidal packed beds to polymer composites. In particular Laval \textit{et al.} present results of how colloidal crystals embedded with gradients and fluorescent barcodes can be fabricated from a dilute dispersion of monodisperse colloids and fluorescent particles. These recent works show that directional drying is a powerful way to build materials with predetermined specifications.}

Despite much progress in the understanding of drying, many questions must still be answered. What is the role of the particle size? Does the solid region remains wet during growth? What is the water concentration in the solid region uniform? What are the local drying dynamics? How do particles organize at the interface? And which factors control the dynamics and ordering of particles? Dynamic, particle-scale observations are required to answer some of these questions and in the following we will answer to a few of them.

The organisation of particles is almost always observed by looking at the dried sample, with optical,~\cite{goehring_solidification_2010} confocal,~\cite{Bodi:2010,Mont:2011} atomic force,~\cite{piroird_role_2016} or electron microscopy.~\cite{inasawa_formation_2016} The particle velocities were previously measured using markers (large particles), which is not ideal~\cite{Bodi:2010} as it only provides an indirect measurement. The dynamics were also assessed indirectly by SANS/SAXS,~\cite{li_drying_2012,Boul:2014} which only provide an average measurement of the particle behaviour.

To progress in the understanding of drying of particle suspensions, we investigate unidirectional drying of weakly Brownian particle suspensions by laser scanning confocal microscopy.~\cite{Yosh:1991} Using a high imaging rate (up to 20~Hz) we can follow \textit{in situ} individual particles dynamics followed by their ordering at the liquid/solid limit. We correlate the particle dynamics to their short/long range ordering organization and find that faster particles velocities yield better ordering and denser packings.

\section{Material and methods}
\subsection{Sample preparation}
Monodisperse spherical PtBMA/PFEMA particles, 1.1~$\mu$m in radius, as measured by microscopy image, were synthesized by dispersion polymerisation with a final density $\rho=1.16$~g/cm$^3$, and fluorescently tagged with BODIPY 545.~\cite{Kodger:2015} \added{The polydispersity index was about 0.1 as measured by Dynamic Light Scattering.} They were dispersed in 0.1~mM aqueous solution of Sulforhodamine B (Fluotechnik, France), a fluorescent dye whose absorption/emission peaks are at 566/584~nm in water. The use of this secondary fluorescent dye allows us to follow the receding front of the solvent. The particle concentration of the suspension was $\Phi$=10 vol\%. Zeta-potential measurements showed that particles are negatively charged (-30 mV), thus the repulsive force hinders aggregation, and the suspensions are readily and properly dispersed. To ensure an optimal dispersion of particles prior to measurement, the suspensions were sonicated for 1 hour in an ultrasound bath. Because of their density and size, particles settle within a few minutes in the thin sample. The diffusion of particles in the suspension is very limited, thus we define them as weakly-Brownian particles. 

\begin{figure}[h]
\includegraphics[width=0.5\textwidth]{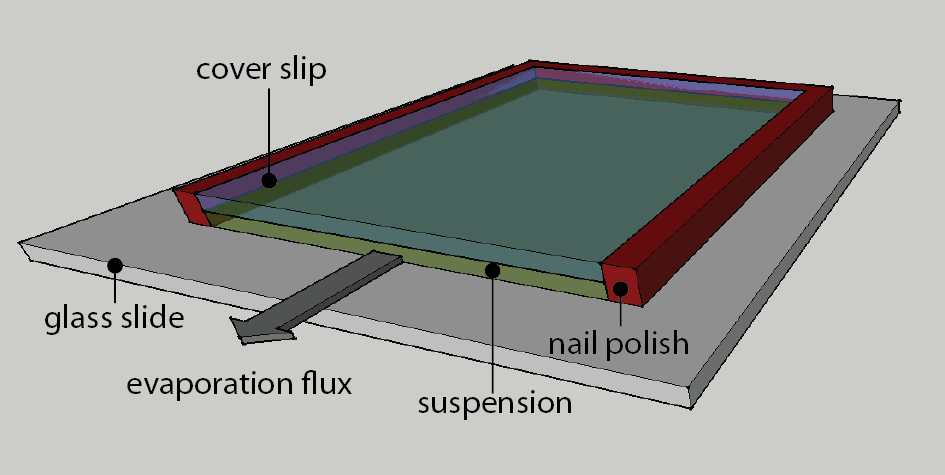}
\caption{Sketch of the Hele-Shaw cell with the suspension inside. Three sides are sealed, forcing unidirectional drying on the opened side.}
\label{fig:scheme}
\end{figure}

To observe the drying process, thin Hele-Shaw cells were prepared as follows: a 8~$\mu$L droplet of suspension was placed onto a microscope glass slide (VWR). A square cover slip (2$\times$2~cm$^2$ and thickness $\approx$~170~$\mu$m, VWR) was carefully deposited on top of the droplet. The Hele-Shaw cell was sealed on three sides with nail polish. One side of the cell is opened, inducing directional drying (Figure~\ref{fig:scheme}). \replaced{Within the series of measured samples, the apparent as-prepared thicknesses of all the Hele-Shaw cells were in the 10--20~$\mu$m range. The thickness within the single Hele-Shaw cell was uniform as we measured the cell thickness in different points of each cell. These values were measured by microscopic images of the vertical cross-section of the samples.}{The apparent as-prepared thickness of the samples, measured from the microscopic images of the vertical cross-sections, were in the 10--20~$\mu$m range.} Drying occurs at ambient air condition without any control of humidity and temperature. \added{The typical drying time of an entire cell is of several hours, but the formation of the dense zone at the open-end takes about 30 minutes. When air starts to invade the cell and the dense zone stops growing, we observe the formation of labyrinthine structures followed by the appearance of cracks (such events will not be discussed here).}

\subsection{Confocal microscopy}

To follow \textit{in situ} and in real-time the dynamic and organization of particles during drying, we used a Leica TCS SP8 (Leica-Microsystems, Mannheim, Germany) confocal laser scanning microscope. We used an external continuous laser source at 488~nm (blue) to scan the sample and a Leica HCX PL APO CS $20\times$ dry objective. The light is collected by a photomultiplier.

The 488~nm laser beam excites the dye (BODIPY) within the particles. The fluorescence is integrated in the 495-520~nm range. A high-speed scanner allows to record up to 20 frames per second at a 512$\times$512~pixels$^2$ resolution, equivalent to 290$\times$290~$\mu$m$^2$ at the optical magnification 2$\times$. The microscope is controlled in remote by the proprietary Leica LAS-AF software. Due to the time needed to prepare the sample and launch the acquisition, the first images were taken approximately one minute after sealing the sample.	

\subsection{Image analysis}

The collected series of images were post-processed with Fiji.~\cite{schindelin_fiji:_2012} The only digital manipulation, before any further analysis, was a linear contrast adjustment to distribute the signal in the entire dynamic range. The images were then converted to tiff files for numerical analysis. The velocity of the solid/liquid interface was determined by automated measurement of the interface position, based on the sudden change of fluorescent signal at the interface (Figure~\ref{fig:alternative order}). The velocities of particles was measured using the TrackPy Python package.~\cite{dan_allan_trackpy:_2014} We also used a Python code to compute the radial distribution functions, based on the analysis of the particles position obtained with TrackPy.

\section{Results and Discussion}

We first describe the sequence of drying events, and how a dense region is formed and grows before air invasion. We then identify the existence of regions with different ordering, and measure the particle dynamics during drying.

\subsection{Drying events}
In confined and unidirectional drying, the flow of water towards the open-end of the cell carries particles towards the pinned air/water interface where evaporation occurs. Particles progressively accumulate, building a dense region. \replaced{We observe a time-delay (on the order of half a hour) between the onset of drying of the dense zone and the dense zone formation, that remains wet as the water flows through it. This phenomenon was already observed by Divry \textit{et al.}~\cite{divry_drying_2016} in drying a suspension of latex hard spheres: as these authors, we define the observation of crack as the onset of complete drying.}{A similar delay between dense zone formation and drying was already observed by Divry \textit{et al.}~\cite{divry_drying_2016} in drying a suspension of latex hard spheres.}

\replaced{In particular, while the air/water interface is pinned in constant position, at the open-end of the sample, our observations reveal that the dense zone remains wet as it grows: water flows through it, dragging always more particles. As water flows out of the cell, the same volume should be replaced by, for example, air bubbles or change in thickness: we cannot observe these two phenomena because they would appear far from the point where we observe the flux of particles and the growth of the dense zone. }{Our observations reveal that the dense region remains wet as it grows: water flows through the dense region, bringing always more particles, but the air/water interface is pinned in constant position, at the open-end of the sample. In our measurements we do not clearly observe, due to the experimental constraints, bubbles formation along the sealed edge or thickness changes far from the dense domain.}

\begin{figure*}[htbp]
\begin{overpic}[angle=-0,width=11cm]{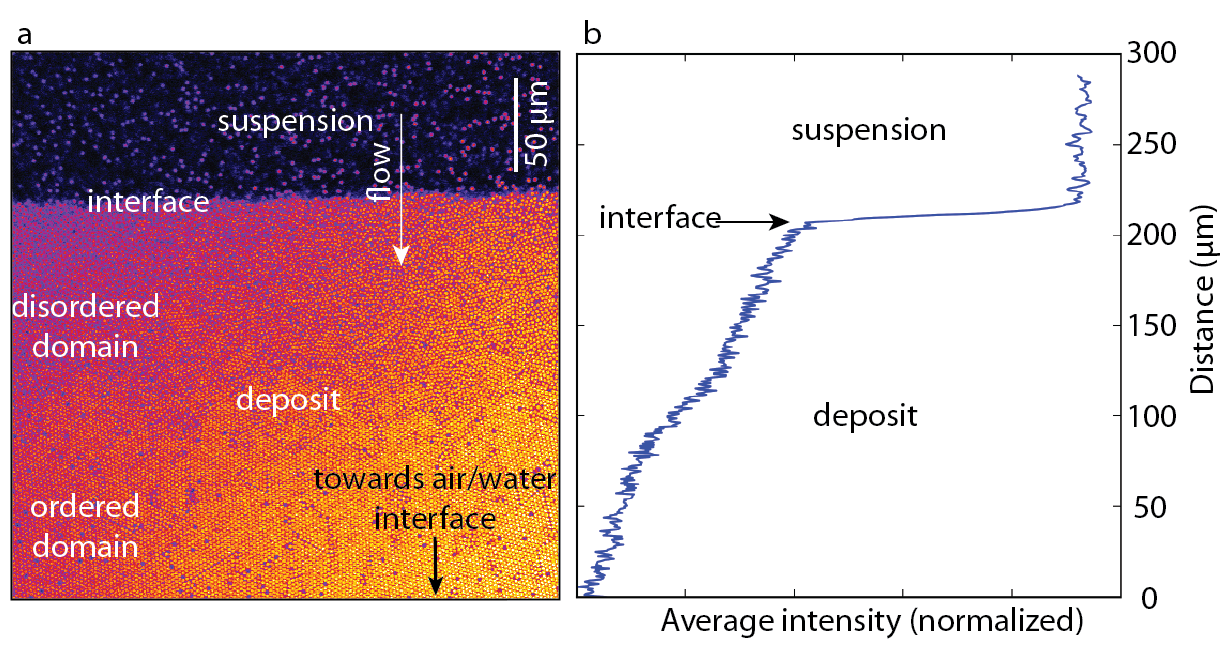}
\end{overpic}
\caption{Alternating ordered and disordered domains observed during the formation of the dense zone, close to the opened side of the cell. a) The suspension flows from the top to the bottom toward the air/water interface which is not shown in the picture. Although some local reorganization of the particle can occur in the dense zone (described in the text), the ordering of the particles is retained during drying. In this case, the dense and disordered domains observed close to the interface between the dense deposit and the suspension will not turn into a colloidal crystal after further drying. \added{The integrated intensity changes along the horizontal and vertical lines as described in the text.} b) The interface between the dense deposit and the suspension is very sharp. We detect this sharp increase to track the interface position and measure its velocity. \deleted{The steady decrease of intensity in the dense deposit, on the plot, is due to a small tilt of the sample.}}
\label{fig:alternative order}
\end{figure*}

Particles, driven by the flow, accumulate in close-contact into a dense region (Figure~\ref{fig:alternative order}(a)). The interface between the dense region and suspension is sharp (Figure~\ref{fig:alternative order}(b)). The particle concentration goes from the nominal suspension concentration to a nearly dense packing over a distance of only 2-3 particles. The diffusion of particles in the suspension is very limited compared to the induced advection, which is why we defined them weakly-Brownian.

We can already observe in Figure~\ref{fig:alternative order}(a) that the particle organisation in the dense region is not constant: ordered (hexagonal) and disordered domains can be observed. \replaced{We outline that the horizontal intensity change is an artifact due to a small tilt of the sample. As the thickness of the optical slice we look at is very thin ($\approx 200$~nm), a tiny tilt is sufficient to induce a linear gradient of intensity over the entire image. The discontinuity in intensity, assigned to the change in ordering, is instead visible through the vertical lines, which are averaged to yield Figure~\ref{fig:alternative order}(b). Albeit it is subtle, it is yet visible. Indeed, on Figure~\ref{fig:alternative order}(b), we observe, at about 100~$\mu m$, that the slope changes. We assign this event to the change in ordering. We can rule out that the change in intensity is an artifact of illumination on the sample because we operated the microscope in confocal mode. }{The imaged intensity of the dense deposit steadily decrease due to a small tilt of the sample. Nevertheless, this decrease could also arise from a change in volume fraction, which does also occur in the deposit. While there is definitely a tilt in the sample and this contributes the most to the decrease, there is a also an small discontinuity in the decrease that correspond exactly where, in the image, the deposit goes from ordered to disordered. It is likely that this change in the average intensity also reflects the change in order, not just the tilt of the sample.}

\begin{figure*}[htbp]
\includegraphics[width=1.0\textwidth]{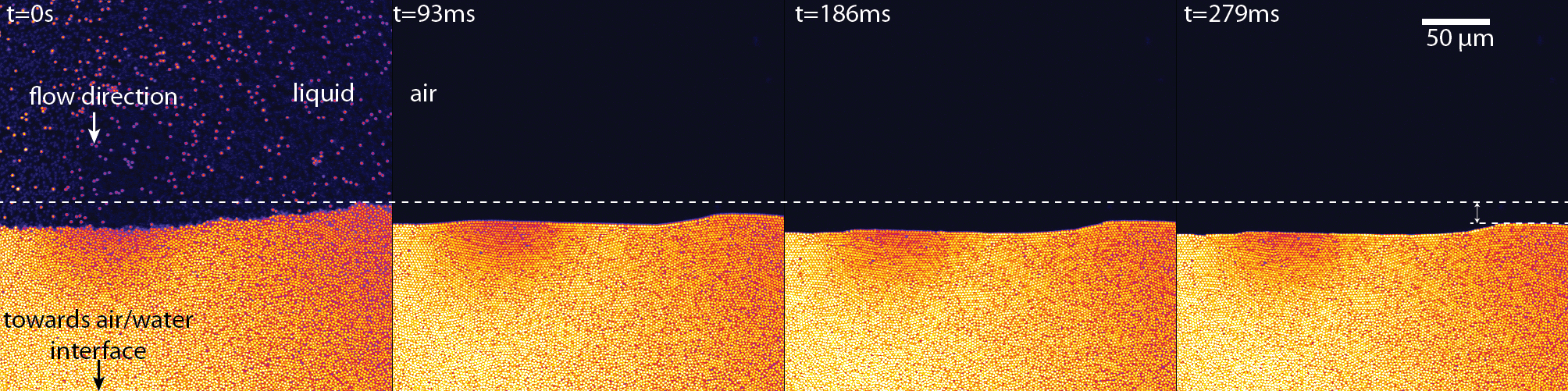}
\caption{Close-up view of the edge of the dense region during air invasion. In the first frame, water still flows through the dense region (top to bottom), towards the open-end of the sample. The air invasion occurs very rapidly. The capillary forces created during the air invasion compact the structure, whose effect is a higher degree of particle ordering. The dashed line is a guide for the eye to follow the recession of the dense deposit. This behaviour is only observed at the edge of the dense deposit. Far from the edge, the order or disordered configuration of the particles is not affected by the air invasion.}
\label{fig:air_invasion}
\end{figure*}

The sample is a closed system with limited supply of suspension. At some point of the drying stage, the pressure drop inside the sample is too high: the air invades the dense deposit (Figure~\ref{fig:air_invasion}). Air invasion occurs too rapidly to be captured in the images. The capillary forces developed during the air invasion pull the particles together in the dense deposit. This locally results in an increase of the crystallinity. In particular, we only observed this behaviour at the liquid/solid interface. In the bulk of the dense deposit, particle ordering is not affected by the air invasion. The structure is probably too compact to further densify, unlike at the edge. 

\begin{figure}[htbp]
\includegraphics[width=0.5\textwidth]{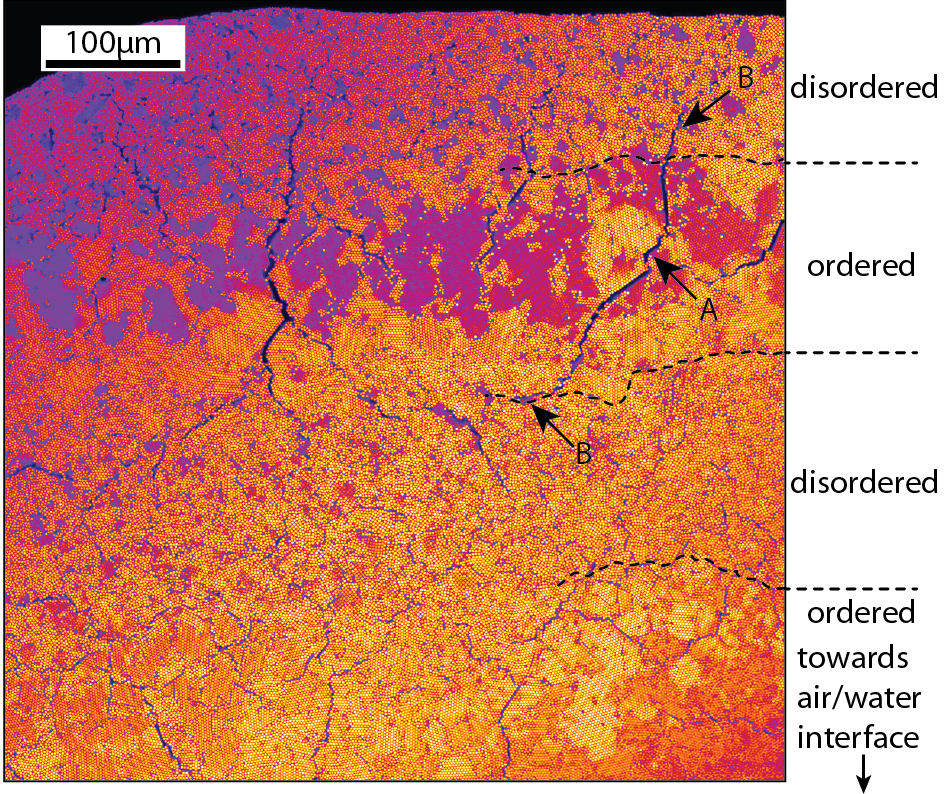}
\caption{Cracks in the dry dense deposit. The formation of cracks starts after air invades the dense deposit. Cracks are preferentially formed in the ordered domains (A) \added{going through the grain boundaries}, they are deflected and stopped in the disordered domains (B).}
\label{fig:cracks}
\end{figure}

The air invasion also marks the onset of crack growth in the dense deposit (Figure~\ref{fig:cracks}). \added{This feature has also been observed during the condensation, by water pervaporation, of charged colloidal particles in microchannels.~\cite{Zian:2015}} Cracks propagate well in ordered domains whereas they are deflected and stopped when they reach the disordered ones. \added{Indeed, in the ordered zone, cracks propagate along the grain boundaries that are defects in the crystal. In the disordered zone, we cannot define neither a defect in the crystal structure nor a grain boundary through which the cracks can start and propagate easily. To make easier to understand the picture, we can say that the cracks become indiscernible in the disordered domains.}

The air invasion is responsible of another drying regime, resulting in a labyrinthine pattern inside the cell,~\cite{Sand:2007,Knud:2008,Sand:2011} which will not be described in the present work.

\subsection{Evidence of alternate order in the dense deposit}

At the end of drying and particles flow, the deposit is made of alternating ordered (hexagonal) and disordered (glassy) domains. In the observation plane, the particle density is apparently greater in the ordered domains than in the disordered ones. 

\begin{figure}[htbp]
\includegraphics[width=0.5\textwidth]{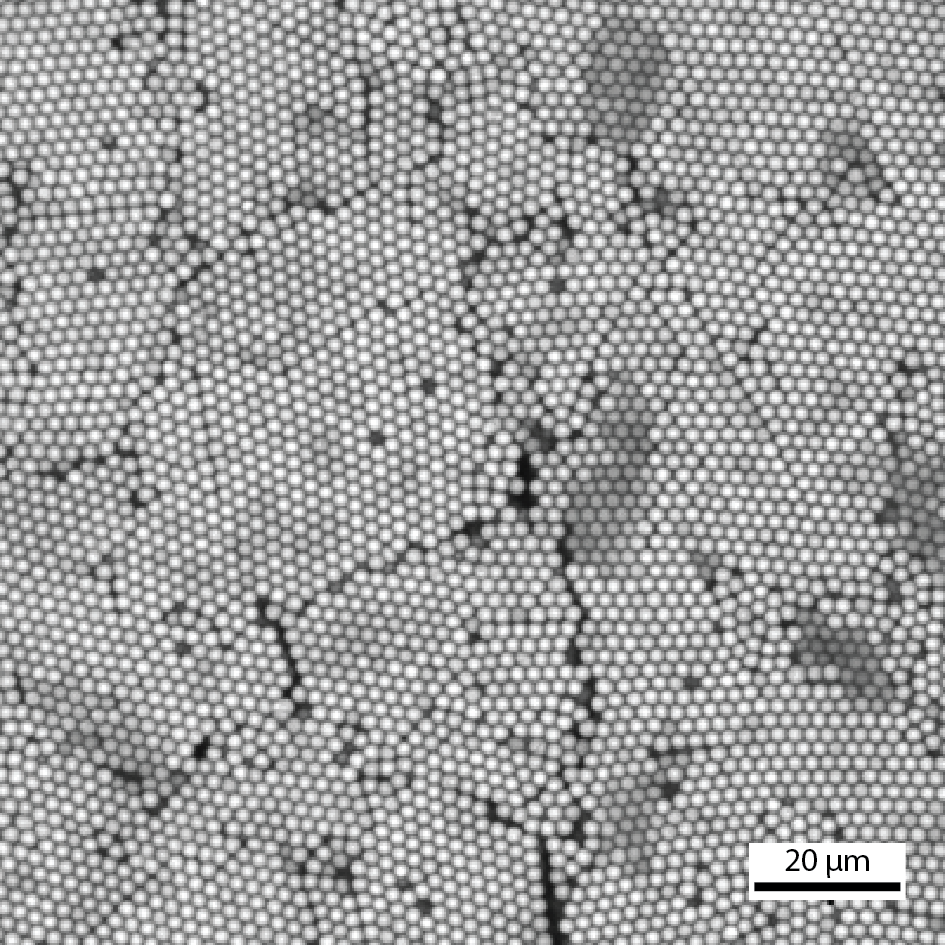}
\caption{Close-up view of \deleted{a typical structure of} an ordered domain in a dried sample (no water flows anymore). The dense deposit is composed of close-packed domains with cracks between ordered domains.}
\label{fig:orderedZone}
\end{figure}

\begin{figure}[htbp]
\includegraphics[width=0.5\textwidth]{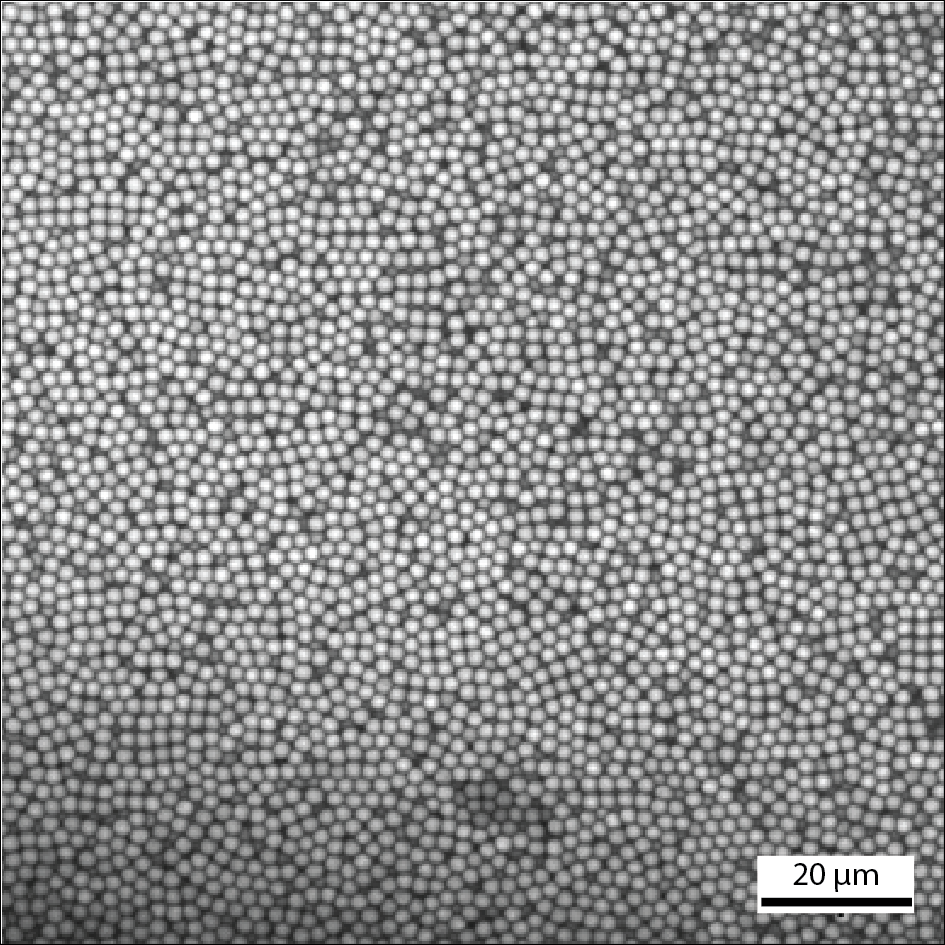}
\caption{Close-up view of \deleted{a typical structure} of a disordered domain in a dried sample (absence of water flow). The structure is dense and disordered, with an apparent short-range order only.}
\label{fig:disorderedZone}
\end{figure}

Close-up views of the particle organization in the ordered and disordered domains are shown in Figures~\ref{fig:orderedZone} and \ref{fig:disorderedZone}. \added{From these two figures, we estimate the 2D density of particles to be 77\% in ordered domains and 62\% in disordered ones.}

The ordered domain, Figure~\ref{fig:orderedZone}, is made of 2-dimensional, long-range ordered, colloidal crystal grains. Defects typical of colloidal crystals can also be observed, such as vacancies, stacking faults, or dislocations. Owing to the well ordered domains, we argue that the ordering should continue in the interior of the domains, in a three-dimensional fashion. This fact was already observed in rather similar experiment by \textit{post-mortem} electron microscope images.~\cite{inasawa_formation_2016}

We observe that the ordered structure is formed \replaced{as soon as the particles reach the dense zone}{instantly}. The particles carried by the flow settle in a crystalline, hexagonal organisation at the solid/liquid interface. No further particles reorganization is observable at the current level of magnification ($40\times$) \added{and resolution}.

\begin{figure}[htbp]
\centering
\includegraphics[width=0.5\textwidth]{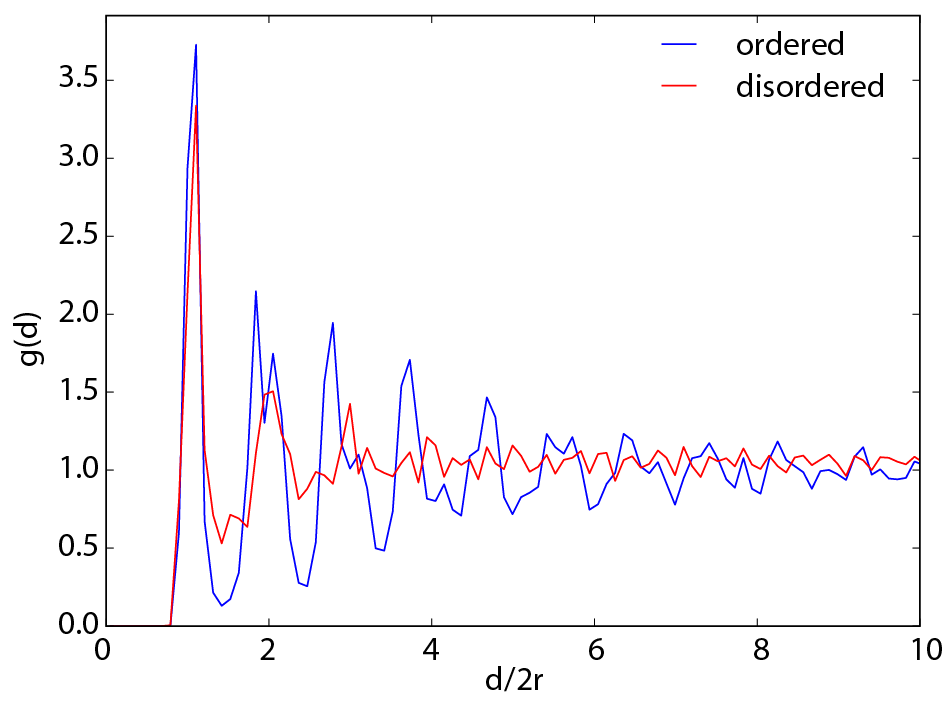}
\caption{Radial distribution function $g(d)$ through the ordered/disordered dense domains: (ordered) long-range quasi-colloidal crystal extends up to several orders of neighbours \textit{versus} (disordered) short-range ordering. $r$ is the radius of particles.} 
\label{fig:gdr}
\end{figure}

The disordered domains, shown in Figure~\ref{fig:disorderedZone}, are free from long-range cracks, \replaced{as it is expected for a disorder colloidal packing. Both disordered and ordered features were also observe by Ziane \textit{et al.}~\cite{Zian:2015} in colloidal crystals growth by pervaporations in microfluidic channels.}{This zone is characterized by short-range order (1 or 2 neighbours), typical of a dense and disordered colloidal packing.} 
To better quantify the particle order in the dense deposit, we computed the radial distribution function $g(d)$, which represents the probability to find a particle at a distance $d$ from a reference one, at several positions close to the cell edge. Typical radial distribution functions $g(d)$ of both phases are shown in Figure~\ref{fig:gdr}.

We find the $g(d)$ confirms, for the ordered case, long-range order extending to several neighbours. The order is limited to the first two neighbours for the disordered case. The $g(d)$ of the ordered domain strengthens the thesis of readily made \replaced{hexagonal }{\emph{hcp}} 2-dimensional lattice (made of several grains), whereas the $g(d)$ for the disordered domain is the expected for a randomly packed structure. From these results we cannot infer the actual numerical value of the particle density \added{that is a measure of the three-dimensional packing}.

We have shown that directional drying causes the formation of hexagonally ordered domains, and large, dense, and disordered domains. The presence of these two different organizations implies that at least two different mechanisms cause the ordered/disordered arrangements. We hypothesize that the particle collision velocity at the solid/liquid interface is the origin of the mechanism of ordering.

\subsection{Dynamic study of the formation of the dense deposit}

The rapid imaging mode of the microscope allows us to individually follow the trajectory of particles during the formation of the dense deposit. From the trajectories we measure individual particles velocities $v_p$ so that we may identify the ordering mechanism during the formation of the dense deposit. $v_p$ was measured with the particle tracking package TrackPy,~\cite{dan_allan_trackpy:_2014} which detects the particles, track their position, and compute their velocity. \added{Particles are tracked from about $60~\mu$m from the interface until they reach the interface. As particles are weakly-Brownian, their velocity directly results from the velocity of the water flow towards the interface. There are therefore little variations in their velocity as the interface is approached.} 

\begin{figure*}[htbp]
\centering
\includegraphics[width=18cm]{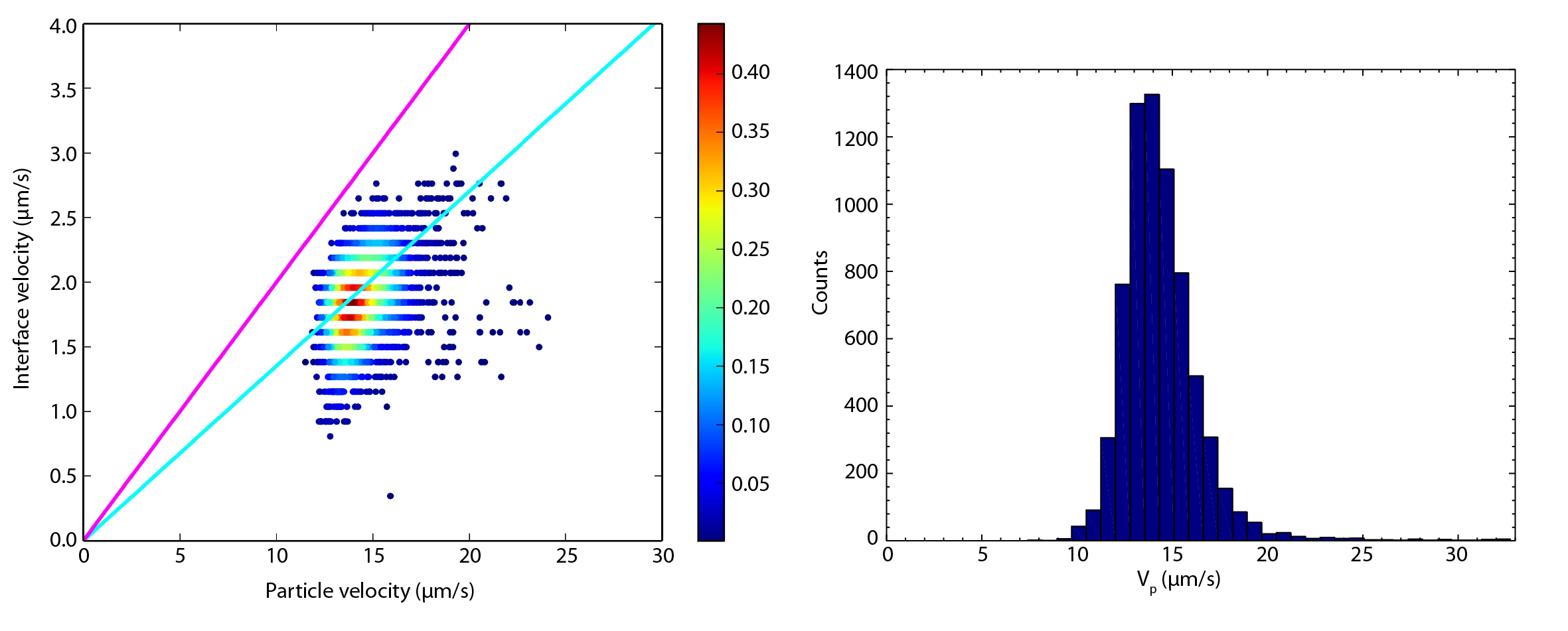}
\caption{Left: Interface velocity $V$ \emph{versus} particle velocity $v_p$, measured from images series where ordered domains are formed. Colour bar represents the percent fraction of the total data. Continuous lines are drawn from the simple mass conservation model, see Figure~\ref{fig:mass conservation}: blue line for ordered zone, $\Phi_d =0.74$, pink for dense and disordered domains, $\Phi_d = 0.5$, see text. Right: histogram of the distribution of $v_p$. The velocity distribution is clearly non-gaussian.}
\label{fig:hexagonal}
\end{figure*}

\begin{figure*}[htbp]
\centering
\includegraphics[width=18cm]{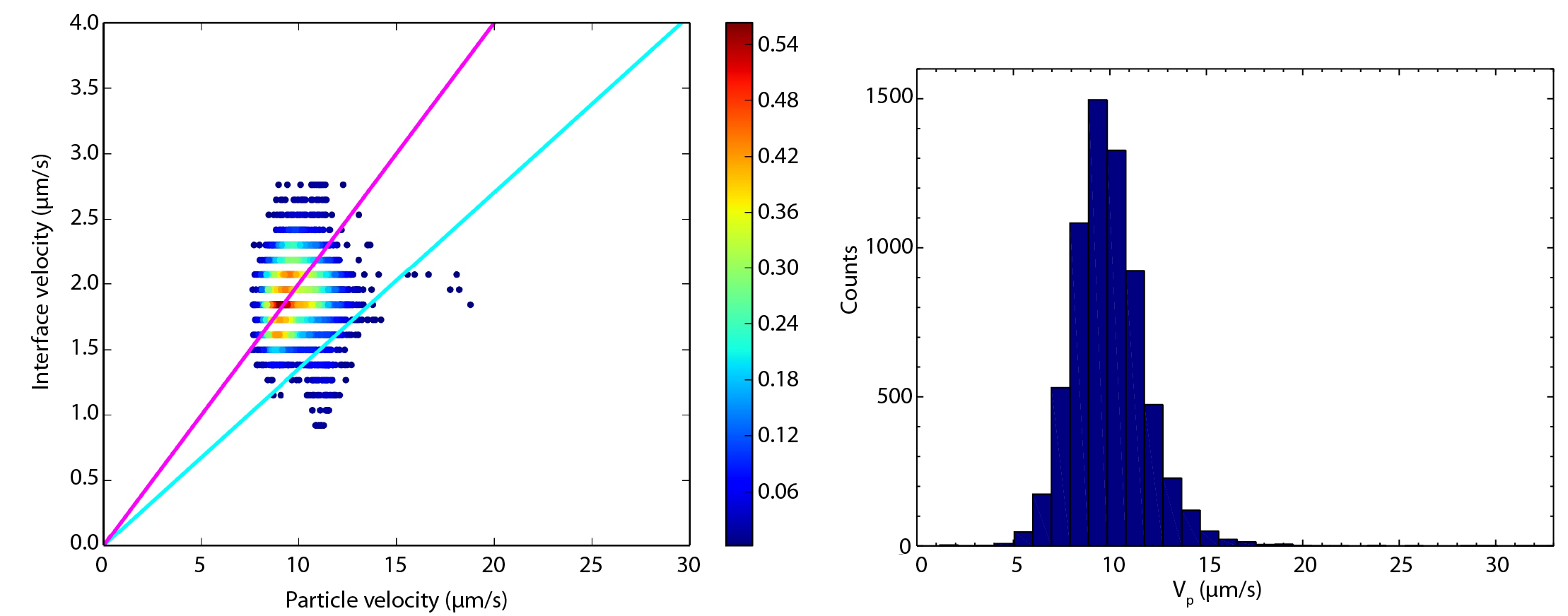}
\caption{Left: Interface velocity $V$ \textit{versus} particle velocity $v_p$, measured from images series where disordered domains are formed. Right: histogram of distribution of $v_p$. Both plots as in Figure~\ref{fig:hexagonal}.}\label{fig:glassy}
\end{figure*}

In the same series of images, we can also determine the interface velocity $V$ of the packed/dilute interface. $V$ is computed by measuring the interface position in each sequence of images. The results of \textit{V versus $v_p$} are shown in Figure~\ref{fig:hexagonal} for the ordered domains (colloidal crystal) and in Figure~\ref{fig:glassy} for the disordered domains. \added{Each point drawn in these two figures is represented by at least hundreds of tracked particles: in particular each red point is represented by a population of thousands of particles.}

\begin{figure}[htbp]
\centering
\includegraphics[width=0.5\textwidth]{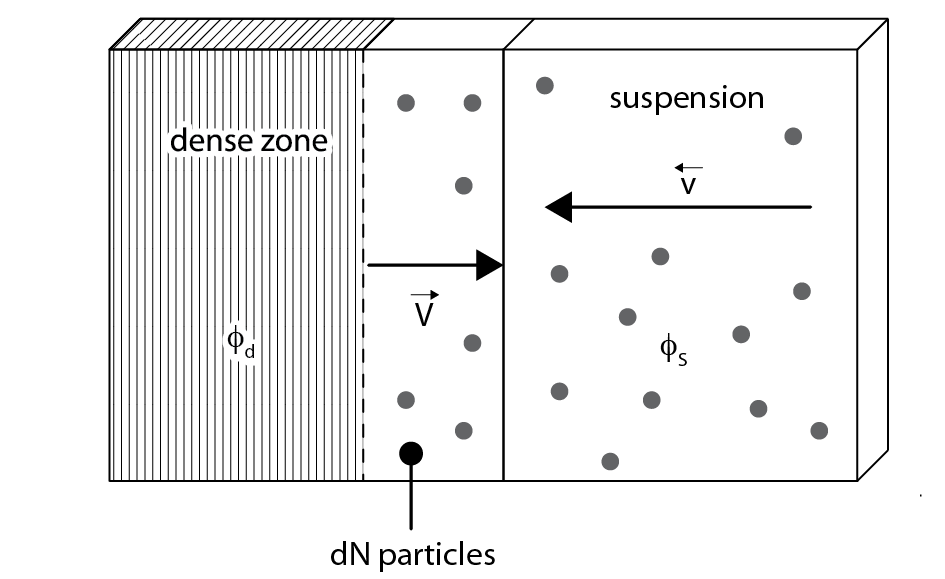}
\caption{Mass conservation model used to draw the continuous lines in Figs.~\ref{fig:hexagonal} and \ref{fig:glassy}. \added{$S$ is the cross-section of the cell.} $\Phi_d$ and $\Phi_s$ are respectively the volume fraction of particles in the dense deposit and in the suspension. $V$ is the velocity of the interface (where the deposit grows) and $v_p$ is the velocity of particles arriving from the suspension.}
\label{fig:mass conservation}
\end{figure}

\replaced{We measured an average $\bar{v}_p=14\pm 2~\mu$m/s for the ordered packing and similarly $\bar{v}_p=9\pm 2~\mu$m/s for the disordered domains. In Figures~\ref{fig:hexagonal} and \ref{fig:glassy} we represented the distributions of $v_p$ that were used to build the $V$ \textit{versus} $v_p$. We can observe that the distribution are clearly non-gaussian.}{We observe the largest fraction of data points are centered at $v_p = 14.2~\mu$m/s for the ordered packing and $v_p \approx 8~\mu$m/s for disordered domains.} The interface velocity $V$ is accidentally the same ($V \approx 2~\mu$m/s) for the two types of ordering. 

To assess a possible correlation between the $V$, $v_p$, and order/disorder we propose a simple mass-conservation model, schematically represented in Figure~\ref{fig:mass conservation}.

We suppose that the number d$N$ of particles in the newly formed region of the dense deposit during the time d$t$ is equal to the number of particles arriving from the suspension at the same time, that is: 	

\begin{equation}
\frac{\mathrm{d}N}{\mathrm{d}t} = V \cdot \Phi_d \cdot S = v_p \cdot \Phi_s \cdot S \Longleftrightarrow V = \frac{\Phi_s}{\Phi_d} \cdot v_p
\label{eq:equation mass conservation}
\end{equation}

where $\Phi_d$ and $\Phi_s$ are respectively the volume fraction of particles in the dense deposit and in the suspension, and $S$ is the cross-section of the cell, \added{see Figure~\ref{fig:mass conservation}.} 

\begin{figure}[h!]
\includegraphics[width=0.5\textwidth]{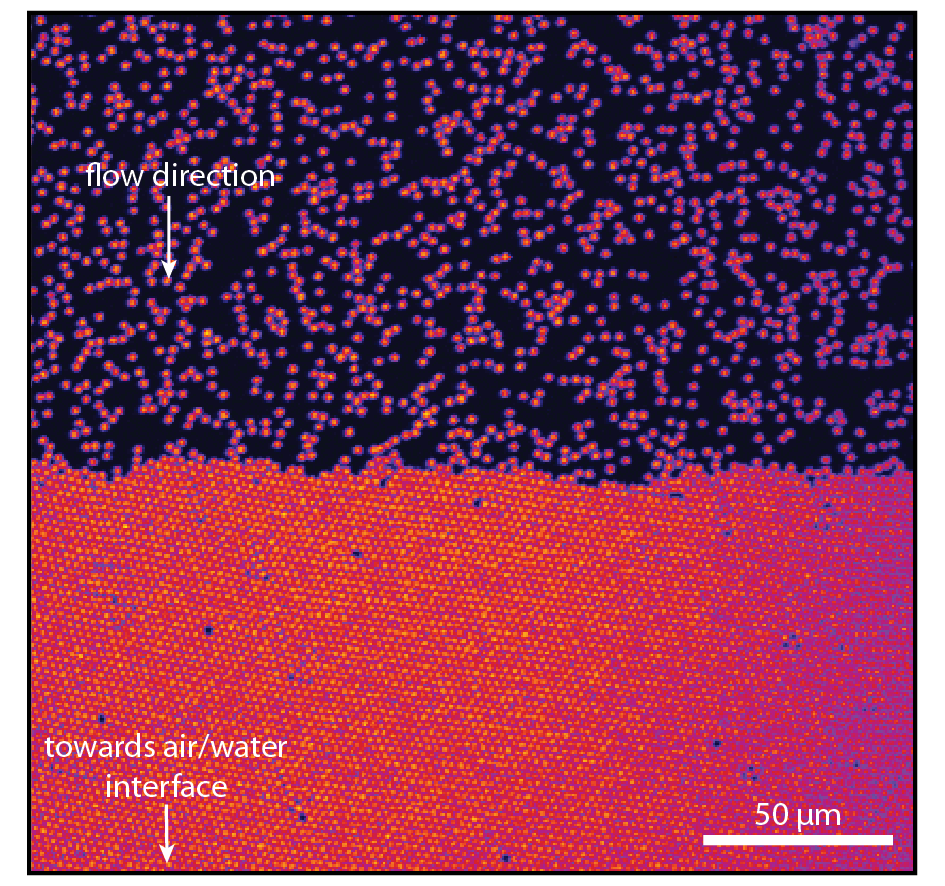}
\caption{Close-up view of the dense deposit/suspension interface during the growth of ordered domains. The ordered configuration of particles is formed as the particles arrived at the deposit interface.}
\label{fig:interfaceHexagonal}
\end{figure}

Using our model with $\Phi_s = 0.1$ in the suspension and $\Phi_d = 0.74$ characteristic of a dense crystalline structure, we have drawn the continuous blue lines in Figures~\ref{fig:hexagonal} and \ref{fig:glassy}. Although we cannot really determine the real three-dimensional structure of the dense deposit (unless by breaking the dry sample to image the transversal section), the 2D-image clearly shows the characteristic pattern of a two-dimensional \emph{hcp} or \emph{fcc} organization, both of which have the same particle density $\Phi=0.74$. The prediction of the model (blue line) crosses the largest fraction of data points (red area) of Figure~\ref{fig:hexagonal}. \added{At the spatial length scale of Figure~\ref{fig:interfaceHexagonal} (one hundred of particles per horizontal line) and time-scale, we observe that the particles, once they have reached the densification front, freeze at the contact point and do not locally reorganize or move.} This remarkable coincidence \replaced{might imply }{implies} that this region forms instantaneously (within a time interval between two frames, which is 0.05~s) as a dense crystalline lattice and it does not become packed afterwards. The precise crystalline nature of the dense packing in 3D is still to identify. 

We draw the pink lines in Figure~\ref{fig:glassy} such that it crosses the area with the largest fraction of points (red areas). Within our model, the volume fraction measured by this qualitative approach is $\Phi_g=0.5$. This $\Phi_g$ value approaches the onset of glassy behaviour ($\Phi_g=0.58$). We note that the phase diagram for hard-sphere particles is quite complex with crystal/liquid coexistance starting at a volume fraction of 0.494~\cite{Schoe:2007} continuing to 0.545, and then from 0.545 to 0.74 being crystalline. Additionally, the glass transition occurs at 0.58 and continues to 0.64 which is generally accepted as randomly close packed. To simplify the phase identification in our visual observations, we will consider a phase diagram consisting of either glass or hexagonal lattice. For clarity, the same pink line is replicated in Figure~\ref{fig:hexagonal}. The difference between the measured particle fraction and the onset of glassy behaviour might be due to a different ordering induced by the glass wall. We note in Figure~\ref{fig:disorderedZone} an unfocused bottom layer of particles that allow us to confirm the three-dimensionality of the lattice. However, due to the dense packing, the imaging depth is limited to two layers of particles and we are unable to investigate the particle organization within the bulk volume. 

The correlation between particles velocities and order is a counter-intuitive observation after the work of Mar\'in \textit{et al.}~\cite{marin_order--disorder_2011} about the drying of water-based suspensions in sessile drops (particle radius in the range of $0.25-1 ~\mu$m). The authors elegantly showed that ordering (\emph{hcp} or \emph{bcc/fcc} colloidal lattice) is caused by particles slowly colliding with the solid/liquid interface. These colloidal particles, due to their Brownian motion, can rearrange into an ordered structures. The colloidal particles that arrive rapidly hamper the rearrangement of the neighbours because they quickly form aggregates.
	
A phenomenon similar to what we observe has been recently described by Piroird \textit{et al.}~\cite{piroird_role_2016} The authors' explanation on how order takes place while drying colloidal silica nanospheres (12~nm radius) in controlled atmosphere, is based on the rate of evaporation of the solvent. For small evaporation rate, particles aggregates form in the suspension that can not order as they reach the interface. During fast drying, the particles have no time to form intermediate aggregates, and impinging singularly onto the solid/liquid interface, they can order in closely packed structures. In our conditions, we never observed any particles aggregation.

We observe that these two reports are apparently in contradiction and predict that particles size is one of the key conditions to observe ordered/disordered particles arrangements in drying experiments. 
Despite the similarities between the reported works and ours, we reiterate that our samples are unidirectionally dried in a confined Hele-Shaw cell with a single open-end. We believe that the formation of quasi-organized structures is more likely due to the energy necessary to reorganize the particles as they arrive at the interface.

The counter intuitive result that we observe is related to particles that are only weakly Brownian. The Peclet number $Pe$, which is the ratio of diffusion time to hydrodynamic time, $Pe=6\pi\eta\frac{v_p R^2}{\mathrm{k_B}T}$ (\replaced{$R=1.1~\mu$m }{$R=1~\mu$m}, radius of the colloidal sphere, $\eta$ the water viscosity) is \replaced{51 }{180} and \replaced{79 }{270} for $v_p$ equal to \replaced{9 }{8} and 14~$\mu$m/s respectively. In our system, the particles do not have time to explore configurations by diffusion as they reach the dense deposit. Their motion is controlled by the hydrodynamic flow which drags them to the open end because of the solvent evaporation. 

The flow of water through the packed particle deposit generates a pressure gradient that can be estimated using Darcy's law and Kozeny-Carman's equation:

\begin{equation}
\frac{\Delta P}{L}= k \mu v_p S_v^2\frac{\phi_f^2}{(1-\phi_f)^3}
\end{equation}

where $L$ is the length of the deposit, $\phi_f$ the particle volume fraction, $\mu$ the viscosity, $S_v=3/R$ the ratio of surface to volume of the particles for ideal spheres, and $k$ a constant whose value is generally 5.~\cite{inasawa_formation_2016}
This equation shows that a higher velocity leads to a higher pressure gradient and consequently to a higher packing of the particles. It was confirmed experimentally by Inasawa \textit{et al.}~\cite{inasawa_formation_2016} that higher impinging velocities lead to higher volume fraction in the deposit in unidirectional packing. Moreover the pressure acting on the particles may help to overcome the electrostatic repulsions between the particles, leading to short range attraction between the particles. Consequently the authors observe a saturation of the volume fraction in the deposit around $\phi=0.5$ corresponding to a disordered deposit as attractive particles cannot rearrange as they reach the deposit.

\begin{figure}[h!]
\includegraphics[width=0.5\textwidth]{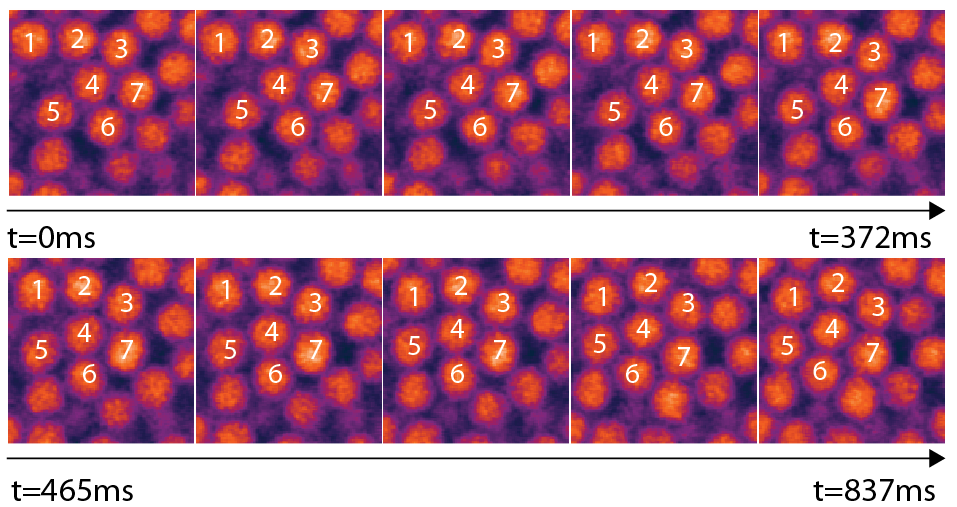}
\caption{Sequence of images showing some reorganization in a disordered domain behind the deposit interface. Time between each frame: 0.093~s. Some particle reorganization is observed, improving the packing density, but not enough to yield an ordered domain. The region shown here is located approximately 20~$\mu$m behind the deposit interface (10 particle layers).}
\label{fig:denseZoneGrowing}
\end{figure}

In our case, we find experimentally $\phi_f=0.74$ and $v_p=14~\mu$m/s in the ordered domains while for the disordered domains $\phi_f=0.5$ and \replaced{$v_p=9~\mu$m/s }{$v_p=8~\mu$m/s}. Taking these two sets of values and $L=10^{-3}$~m, we find $\Delta P=16$~kPa for the ordered domains and $\Delta P=0.6$~kPa for the disordered domains. The stress pushing on each particle $F=2\pi r^2 \Delta P$ is of the order of 120~nN for the ordered domains and \replaced{5~nN}{4~nN} for the disordered domains. \replaced{We then calculate the DLVO potential taking into account the Van der Waals attraction and electrostatic repulsion. We find that the DLVO barrier is of the order of 12~nN. Therefore the Darcy's stress is sufficient to overcome this barrier in the case of the ordered domains but not for the disordered domains. Consistently, very little particle reorganization occurs in the dense region after its formation (Figure~\ref{fig:interfaceHexagonal}) while we show in Figure~\ref{fig:denseZoneGrowing} that once the particles are embedded in the disordered zone they can undergo some Brownian motion but this local motion is insufficient to result in disordered domains rearranging into ordered ones. }{Thus, we believe that this force is not sufficient to induce particles aggregation as we observe the particles fluctuating individually and locally in both the disordered and the ordered domains. For example, a snapshot of the growth of an ordered stripe is shown in Figure~\ref{fig:interfaceHexagonal}. Very little particle reorganization occurs in the dense region after its formation. The observed ordered/disordered structures are instantly formed as particles are incorporated into the dense zone (Figure~\ref{fig:interfaceHexagonal}). In Figure~\ref{fig:denseZoneGrowing} we show that once the particles are at rest in the dense zone they can undergo some Brownian motion even if the Peclet number decreases to zero at rest. However, particles within the deposit only undergo local perturbations at the scale of two or three neighbours, which is insufficient to result in disordered domains
297 rearranging into ordered ones.} Water still flows directionally through the dense region, which limits the free movement of the particles and drags them.

The disordered to ordered settlement transition as the particle velocities increase is similar to what has been observed in the formation of crystals with non-colloidal particles. In these cases, an external source of energy, for example ultrasound or mechanical agitation, is necessary to organize grains in a crystal form.~\cite{pouliquen_crystallization_1997,lash_scaling_2015} Likewise, the pressure gradient on the particles due to the solvent flow toward the open side of the cell can then be considered as the source of energy necessary to induce particle organization.

\section{Conclusions}
The drying of particle suspensions is a complex phenomenon where evaporation rate, difference in chemical potential of solvent and environment, and size of colloidal particles play interconnected roles. From our observations we find that confined drying by directional flow behaves differently to drying of colloidal suspension of sessile drop. \deleted{We have found that, at the same physico-chemical conditions, the particles sizes causes different results.}

By tracking \textit{in-situ} and in real time the evaporation-induced motion of weakly-Brownian particles close to the open end of a Hele-Shaw cell, we found that the structure of the solid deposit is controlled by the velocity of the particles impinging on the pinned dense zone. \replaced{In our actual ambient and experimental conditions, we found two particle velocity regimes that allow us to define a particle velocity threshold, only valid for the current experiment. Particles form disordered domains without aggregation on their way to the dense deposit when travelling below this threshold (slow regime).}{For the current size of particles and experimental conditions, we established that for particles velocities below the threshold of $v_p< 10~\mu$m/s the particles form disordered domains without aggregation on their way to the dense deposit.} Ordered domains, \replaced{for which we cannot infer the actual three-dimensional crystalline nature,}{(\textit{hcp} or \textit{fcc} lattice)} are formed when the particle velocities are above this threshold (fast velocity branch). \added{Albeit we defined this kinetic threshold about $v_p \leq 10~\mu$m/s, we cannot rule out if other factors contribute to the order/disorder organization.} It should be stressed that the velocity of the solid/liquid interface does not significantly change for ordered or disordered accretion. A better control of both the evaporation rate in constrained experimental conditions and the Hele-Shaw cell will give a strong impulse to understand the mechanism of densification by drying of colloidal suspensions with strong benefit in material processing routes.   

\begin{acknowledgments}
The authors acknowledge F. Gobeaux (IRAMIS/NIMBE/LIONS, CEA Saclay) for zeta potential measurements.
\end{acknowledgments}

\section*{Funding}
The research leading to these results has received funding from the European Research Council under the European Community's Seventh Framework Programme (FP7/2007-2013) Grant Agreement no. 278004 (project \emph{FreeCo}). 
\bibliography{biblio-sechage}

\end{document}